# Multi-origin driven giant planar Hall effect in topological antiferromagnet EuAl$_2$Si$_2$ with tunable spin texture

Xiangqi Liu,[1,†] Ziyi Zhu,[2,†] Yixuan Luo,[1,†] Zhengyang Li,[1] Bo Bai,[1,6] Jingcheng Huang,[3] Xia Wang,[4] Chuanying Xi,[5] Li Pi,[5] Guanxiang Du,[3] Leiming Chen,[2,*] Wenbo Wang,[1,6,*] Wei Xia,[1,6,*] and Yanfeng Guo[1,6,*]

[1]State Key Laboratory of Quantum Functional Materials, School of Physical Science and Technology, ShanghaiTech University, Shanghai 201210, China

[2]Henan Key Laboratory of Aeronautical Materials and Application Technology, Zhengzhou University of Aeronautics, Zhengzhou, Henan 450046, China

[3]College of Telecommunication and Information Engineering, Nanjing University of Posts and Telecommunications, Nanjing 210003, China

[4]Analytical Instrumentation Center, School of Physical Science and Technology, ShanghaiTech University, Shanghai 201210, China

[5]Anhui Province Key Laboratory of Condensed Matter Physics at Extreme Conditions, High Magnetic Field Laboratory of the Chinese Academy of Sciences, Hefei, Anhui 230031, China

[6]ShanghaiTech Laboratory for Topological Physics, ShanghaiTech University, Shanghai 201210, China

[†]These authors contributed equally to this work:
Xiangqi Liu, Ziyi Zhu and Yixuan Luo.

Correspondence: W.B.W. (wangwb1@shanghaitech.edu.cn), L.M.C. (lmchen@zua.edu.cn), W.X. (xiawei2@shanghaitech.edu.cn), Y.F.G. (guoyf@shanghaitech.edu.cn)

**Abstract**

**In topological materials, the planar Hall effect (PHE) is often regarded as a hallmark of profound quantum phenomena—most notably the Adler-Bell-Jackiw chiral anomaly and Berry curvature—rendering it an indispensable tool for deciphering the topological essence of emergent phases. In this study, we delve into the PHE and anisotropic magnetoresistance in the recently discovered layered topological antiferromagnet**

**EuAl$_2$Si$_2$. Our analysis of the robust PHE signal (~ 3.8 μΩ cm at 2 K and 8 T) unveils a distinct interplay of mechanisms. While Berry curvature plays a minor role, the dominant contributions stem from classical orbital MR in the field-induced ferromagnetic state and field-suppressed spin fluctuations in the paramagnetic regime. These insights not only position EuAl$_2$Si$_2$—with its highly tunable spin texture—as an exemplary system for probing the intricate coupling between spin configurations and band topology in magnetotransport but also pave the way for designing novel materials with tailored PHE responses, highlighting significant application prospects in quantum sensing, spintronic devices, and topologically protected electronic systems.**

## 1. Introduction

The planar Hall effect (PHE), characterized by a transverse voltage emerging when coplanar electric ($E$) and magnetic ($B$) fields are applied non-parallelly, has become an indispensable probe for investigating topological quantum states and chiral anomaly phenomena in topological phases[1,2]. In Dirac and Weyl semimetals, this effect typically manifests through the Adler-Bell-Jackiw chiral anomaly mechanism, where aligned $E$ and $B$ fields generate chirality-dependent Weyl node population imbalances. This fundamental process yields two key experimental signatures including the distinctive negative longitudinal magnetoresistance ($n$-MR) and a characteristic sin2$\varphi$ angular dependence in PHE measurements[3]. Beyond fundamental studies, the PHE has promising applications in high-sensitivity magnetic field sensing, non-volatile memory devices, and spintronic components due to its strong response to in-plane spin configurations and magnetic anisotropies. However, the interpretation of PHE as an unambiguous topological indicator is frequently complicated by competing conventional mechanisms, including Fermi surface anisotropy-induced orbital MR and spin-dependent scattering effects in magnetic systems[4-7]. A system for clarifying these different mechanisms in generating the PHE remains very rare.

The antiferromagnetic (AFM) compound EuAl$_2$Si$_2$ presents a uniquely instructive platform for elucidating these complex interactions. This remarkable material undergoes a field-driven transition from an AFM axion insulator ground state to a ferromagnetic (FM) Weyl semimetal phase, exhibiting extraordinary transport properties. Most notably, it demonstrates a colossal anomalous Hall effect (AHE) with record-breaking extrinsic Hall conductivity (~ 1.51×10$^4$ S cm$^{-1}$ at 2 K and 1.2 T) that dwarfs its intrinsic Berry curvature contribution. Through magnetic force microscopy (MFM) studies, this exceptional behavior was ascribed to field-tunable domain wall (DW) configurations - with DW spacing exhibiting near-monotonic compression from 975 nm at 0 T to 232 nm at 4 T. The resulting periodic stripe-like DW structures create pronounced skew scattering effects, dramatically enhanced by the proximity of Weyl points (WPs) to the Fermi level ($E_F$)[8].

In this work, we present a systematic investigation of PHE and anisotropic MR across the AFM-FM-paramagnetic (PM) phase transitions in EuAl$_2$Si$_2$. While our measurements reveal a strong PHE signal displaying the conventional $\sin\theta\cos\theta$ angular dependence, comprehensive analysis establishes its primary origin in classical orbital MR (FM state) and field-suppressed spin fluctuations (PM regime) rather than topological chiral anomaly. This conclusion is robustly supported by two key observations including the characteristic "shock-wave" patterns in parametric $\rho_{xx}$-$\rho_{xy}$ plots, hallmarks of conventional orbital effects, and the complete absence of $n$-MR in the FM phase. These findings highlight the critical importance of discriminating between topological and conventional mechanisms in materials where magnetic order and multiband transport coexist. Nevertheless, the rich phase diagram of EuAl$_2$Si$_2$ - spanning AFM, FM, and PM states - establishes it as an ideal model system for exploring the intricate interplay between spin textures, electronic band topology, and emergent magnetotransport phenomena.

## 2. Results and discussion

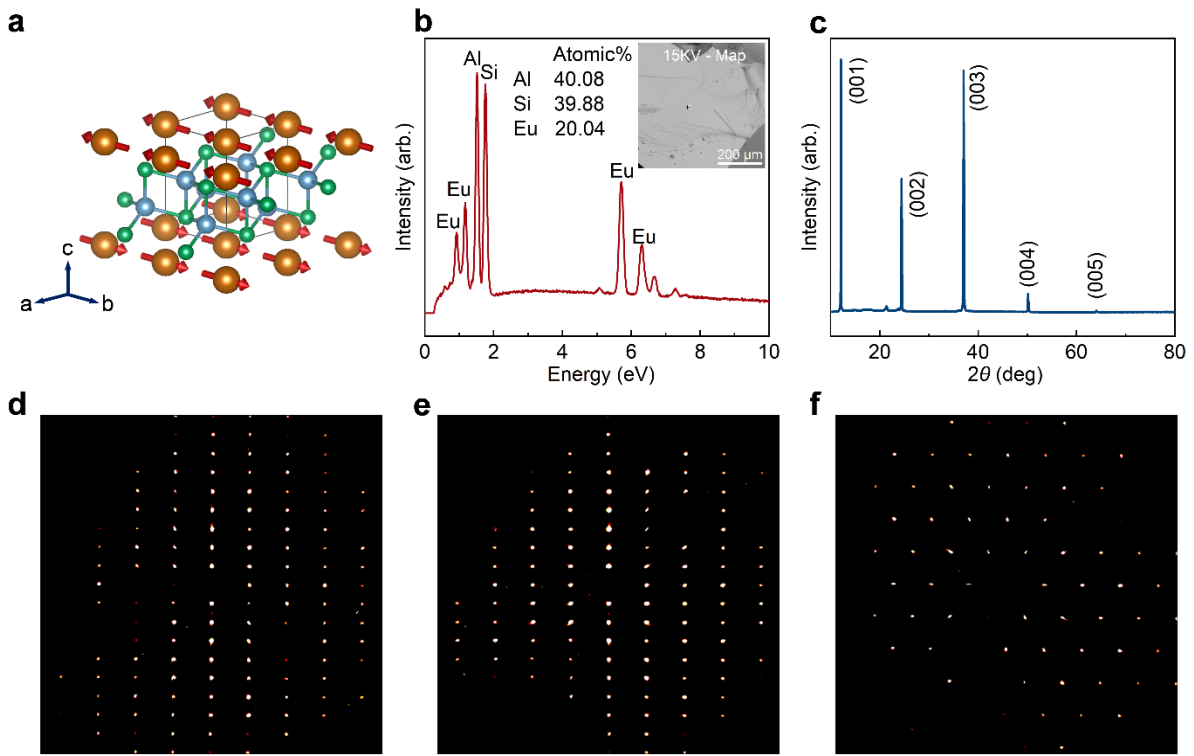

**Figure 1. Compositional and structural characterization of EuAl$_2$Si$_2$.** (a) Crystal structure of EuAl$_2$Si$_2$, showcasing its A-type AFM order characterized by parallel intralayer and antiparallel interlayer spin alignments (denoted by arrows). The structure features two interlaced zigzag chains of Al and Si atoms, sandwiched between Eu layers along the *c*-axis. (b) EDS spectrum acquired in point-scan mode at the location marked by a cross in the inset. (c) Room-temperature PXRD pattern of EuAl$_2$Si$_2$ single crystals, measured along the (00*l*) lattice planes. (d–f) SXRD patterns collected in reciprocal space along the (0*kl*), (*h*0*l*), and (*hk*0) planes, respectively.

EuAl$_2$Si$_2$ crystallizes in a trigonal structure (space group $P\bar{3}m1$, No. 164, **Figure 1a**), as confirmed by powder X-ray diffarction (PXRD) and single-crystal X-ray diffacrtion (SXRD) chracterizations. Refinement of SXRD patterns yielded lattice parameters $a = b = 4.178$ Å, $c = 7.249$ Å, and angles $\alpha = \beta = 90°$, $\gamma = 120°$, with sharp reciprocal lattice patterns (**Figure 1c-f**) attesting to high crystallinity. Energy dispersive spectroscopy (EDS) chracterizations revealed a stoichiometric 1 : 2 : 2 atomic ratio of Eu : Al : Si (**Figure 1b**), further corroborating sample quality. Below the Néel temperature ($T_N \sim 33.6$ K), EuAl$_2$Si$_2$ adopts an A-type AFM spin configuration, characterized by intralayer FM alignment and interlayer antiparallel coupling of Eu$^{2+}$ spins along the *b*-axis (**Figure 1a**). This magnetic transition is marked by distinct signatures in both magnetic susceptibility ($\chi$) and resistivity ($\rho_{xx}$) (**Figure 2a**). The $\rho_{xx}(T)$ curve exhibits metallic behavior ($RRR \sim 62$) above 100 K, followed by a peak at $T_N$ and a sharp drop at lower temperatures, reflecting the opening of a gap in the AFM state. Meanwhile, $\chi(T)$ transitions from isotropic paramagnetism to anisotropic behavior below $T_N$, consistent with spin alignment along *b*-axis.

Our previous first-principles calculations suggest that the AFM state of EuAl$_2$Si$_2$ hosts an axion insulator phase with minute gaps above $E_F$ due to $C_{3z}$ symmetry breaking[8–11]. Intriguingly, an external magnetic field can suppress AFM order, polarizing Eu$^{2+}$ spins to induce a spin-polarized FM state. This transition—governed by the critical saturation field $B_{sat}$ = 4.9 T ($B \,//\, c$-plane, 2 K)—shifts the system to the type-III magnetic space group $P\bar{3}m'1$, breaking time-reversal symmetry while preserving inversion symmetry. Theoretical studies reveal WPs along Γ–A in the Brillouin zone, arising from linear crossings of top valence bands[8], offering a platform for probing topological transport.

The isothermal magnetizations of EuAl$_2$Si$_2$ at 2 K, 50 K and 100 K with the magnetic field along *ab*-plane and *c*-axis are shown in **Figure 2b**. The different saturation fields $B_{sat}$ along *ab*-plane and *c*-axis reveal the anisotropic magnetizations. The saturation moment (~7 $\mu_B$/f.u., matching Eu$^{2+}$ theory) and field-dependent $M(B)$ nonlinearities further underscore the complex magnetic landscape tunable by external stimuli. **Figure 2c-e** show the Magnetic Force Microscopy (MFM) images under magnetic fields of 0.5 T, 3.8 T, and 4.7 T, respectively, with the magnetic field perpendicular to the *ab*-plane. It can be clearly observed that striped magnetic domains exist under low magnetic fields; as the magnetic field increases, the striped magnetic domains gradually become denser. When the magnetic field increases to 4.7 T, the spins are fully polarized, the schematic diagrams of the spin structures corresponding to the above states are shown in **Figure 2f-h**.

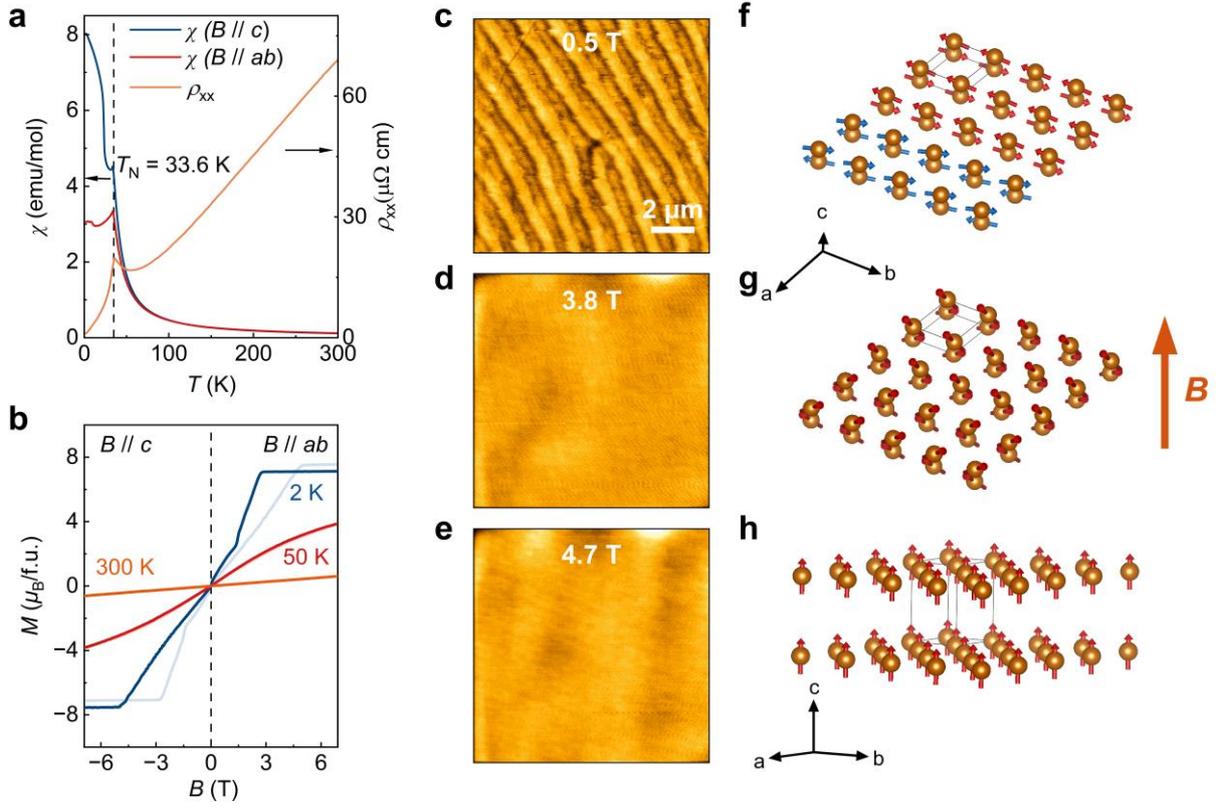

**Figure 2. Magnetic properties of EuAl$_2$Si$_2$ single crystals.** (a) Temperature-dependent magnetic susceptibility $\chi(T)$ measured at 1000 Oe (left axis), and zero-field electrical resistivity $\rho(T)$ from 2 to 300 K (right axis). The dashed line indicates the Néel temperature $T_N \sim 33.6$ K. (b) Isothermal magnetization curves measured below and above $T_N$. At 2 K, the magnetization saturates at $B_{sat} = 2.8$ T for $B // ab$ (left) and 4.9 T for $B // c$ (right). Unsaturated behavior is observed at elevated temperatures due to enhanced thermal fluctuations. (c–e) Sequential MFM images acquired at 6 K under out-of-plane magnetic fields of 0 T, 3.8 T, and 4.7 T, respectively. All scale bars represent 2 $\mu$m. (f–h) Schematic illustrations of the corresponding spin textures under each magnetic field condition.

Magnetoresistance (MR) measurements (**Figure 3b**) under $B \mathbin{/\mkern-3mu/} ab$-plane reveal a rich interplay between spin texture and charge transport. At 2 K, MR transitions from negative (low field) to positive (above 1 kOe), with inflection points mirroring spin-flop transitions in magnetization. Above $T_N$, negative MR emerges from field suppression of thermal spin fluctuations, which reduces carrier scattering[12, 13].

Combined theoretical calculations and angle-resolved photoemission spectroscopy (ARPES) measurements have unequivocally established the existence of linear band crossings near the $E_F$ in EuAl$_2$Si$_2$[8,14]. In the FM state, our first-principles calculations reveal WPs positioned approximately 57 meV below $E_F$ (**Figure 3e**). The topological nature of the electronic structure near $E_F$ is further corroborated by quantum oscillation measurements.

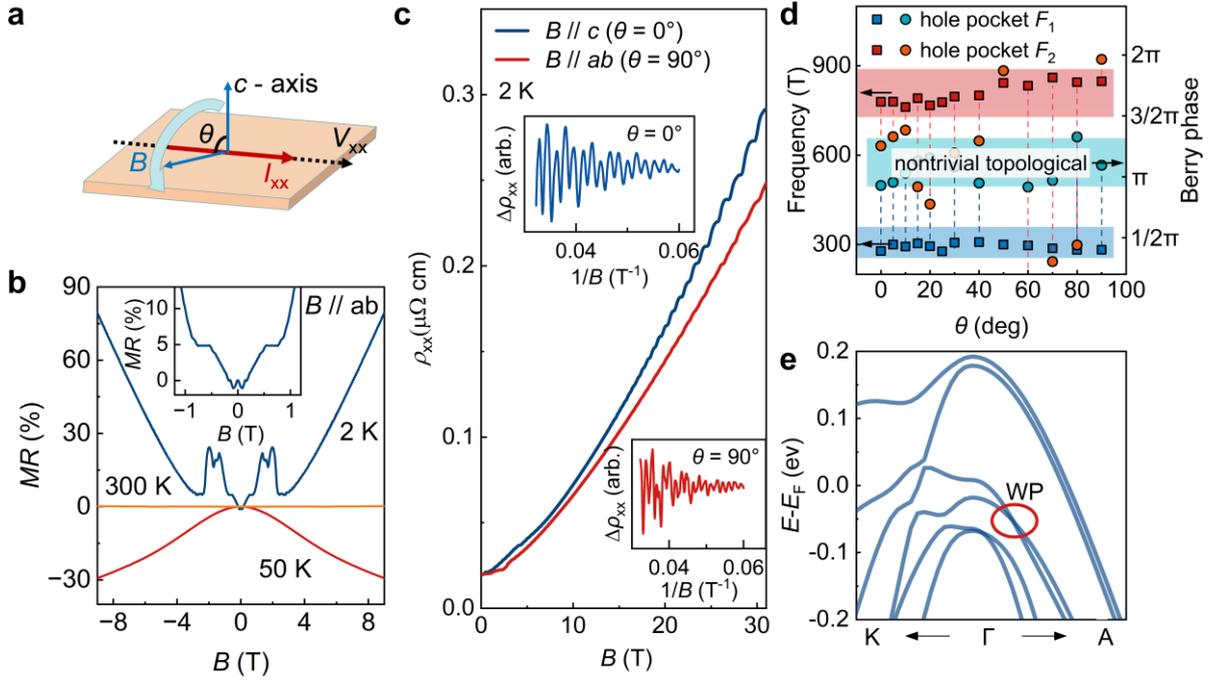

**Figure 3. Electronic transport and topological properties of EuAl$_2$Si$_2$.** (a) Schematic of the measurement geometry, with the magnetic field rotated from the $c$-axis to the $ab$-plane while simultaneously recording both longitudinal voltage $V_{xx}$ and current $I_{xx}$. (b) MR under in-plane field ($B \mathbin{/\mkern-3mu/} b$, $I \mathbin{/\mkern-3mu/} b$) at various temperatures. The inset highlights a weak negative MR below 1000 Oe at 2 K, associated with spin-flop transitions. (c) High-field MR (up to 31 T) at 2 K with field rotation from $B \mathbin{/\mkern-3mu/} c$ to $B \mathbin{/\mkern-3mu/} ab$, reproduced from Ref. [8]. The inset displays the extracted SdH oscillatory components as a function of $1/B$ for the corresponding field directions. (d) Angular dependence of quantum oscillation frequencies ($F_1$, $F_2$) and the Berry phase,

indicating nontrivial topology for the $F_1$ Fermi pocket. (e) Calculated band structure along the K–Γ–A high-symmetry path from Ref. [8], with red circles highlighting the WPs near $E_F$.

**Figure 3c** presents the MR at $T = 2$ K under ultrahigh magnetic fields up to 31 T. Clear Shubnikov-de Haas (SdH) oscillations emerge above 15 T, which were analyzed through fast Fourier transform (FFT) following polynomial background subtraction. This analysis reveals two well-defined fundamental frequencies: $F_1 = 206$ T and $F_2 = 577$ T—which, via the Onsager relation $F = (\hbar/2\pi e)A$, translate to extremal Fermi-surface cross-sectional areas of $A_1 = 1.96$ nm$^{-2}$ and $A_2 = 5.49$ nm$^{-2}$, respectively, establishing a direct proportionality between oscillation frequency and the associated Fermi-surface area. To probe the topological characteristics, we constructed a Landau level (LL) index fan diagram[15-17]. The oscillation peaks and valleys respectively correspond to half-integer and integer Landau indices, with linear fitting of the intercepts revealing Berry phases of $\phi_1 = 0.928\pi$ and $\phi_2 = 1.254\pi$ for the two pockets. First-principles calculations identify $F_1$ as originating from a non-trivial topological hole pocket, while $F_2$ stems from a trivial band. Notably, these calculations are limited to the FM state with $c$-axis spin polarization, leaving the in-plane polarized FM state topology unexplored.

To investigate the angular dependence of band topology, we systematically rotated the magnetic field from the $c$-axis to the $ab$-plane ($\theta = 0°\to 90°$). Remarkably, SdH oscillations persist throughout this angular range, with frequency and Berry phase evolution shown in **Figure 3d** inset. While $F_2$ exhibits slight angular dependence, $F_1$ remains constant - this distinctive behavior confirms the three-dimensional nature of the hole pocket and its robust non-trivial topology across different FM polarization directions[18]. These findings suggest promising opportunities for observing chiral anomaly effects using in-plane magnetic fields.

Following our experimental confirmation of topological band structures in the spin-polarized state, we now explore the predicted signatures of the chiral anomaly through the PHE. Under a strong magnetic field $B$, the Weyl fermions in EuAl$_2$Si$_2$ undergo Landau quantization, forming chiral lowest LLs with their group velocity $v$ locked parallel to $B$. When an in-plane electric field $E$ is applied parallel to $B$, the resulting chiral symmetry breaking leads to interconversion between left- and right-handed Weyl states, producing an anomalous chiral charge imbalance. This quantum phenomenon—the chiral anomaly—is experimentally observable as a pronounced $n$-MR[19–22], serving as a hallmark of nontrivial band topology in Weyl semimetals.

The chiral anomaly manifests distinctly in both planar AMR and PHE when rotating the magnetic field within the $xy$ plane. As depicted in **Figure 4a**, our measurement configuration defines $\theta$ as the angle between the applied magnetic field and current direction. This geometry reveals a complex interplay of electronic transport phenomena, where the raw PHE signal inherently combines contributions from both conventional Hall effects and the topological planar Hall response. Crucially, these components exhibit fundamentally different symmetry properties-the normal Hall resistance demonstrates antisymmetric behavior under field reversal, while the topological PHE signal, arising from chiral anomaly effects, maintains symmetry[23].

To disentangle these contributions, we employ a sophisticated signal processing protocol. First, we eliminate the antisymmetric Hall component through field symmetrization:

$$\rho_{xy}^{sym} = [\rho_{xy}(+B,\theta) + \rho_{xy}(-B,\theta)]/2, \tag{1}$$

where $\rho_{xy} = V_{xy}/I_{xx}$ represents the conventional Hall resistivity. However, experimental realities introduce additional complexity - probe misalignments can admix longitudinal AMR signals into the Hall measurements[7,24,25]. This systematic effect reveals itself through characteristic nodes at $n\pi/2$ intervals in $\rho_{xy}^{PHE}$, betraying the dominant $\cos2\theta$ dependence of in-plane AMR. Our solution involves a secondary antisymmetrization:

$$\rho_{xy}^{PHE} = [\rho_{xy}^{sym}(\theta) - \rho_{xy}^{sym}(\pi-\theta)]/2. \tag{2}$$

This dual-symmetry analysis extends analogously to AMR extraction:

$$\rho_{xx}^{sym} = [\rho_{xx}(+B,\theta) + \rho_{xx}(-B,\theta)]/2 \text{ and } \rho_{xx}^{AMR} = [\rho_{xx}^{sym}(\theta) + \rho_{xx}^{sym}(\pi-\theta)]/2, \tag{3}$$

yielding the AMR ratio:

$$\text{AMR ratio} = [\rho_{xx}^{AMR}(\theta) - \rho_{xx}^{AMR}(\pi/2)]/\rho_{xx}^{AMR}(\pi/2) \times 100\%. \tag{4}$$

Having established this rigorous signal purification framework, we can now quantitatively describe the influence from chiral anomaly through the resistivity tensor:

$$\rho_{xy}^{PHE} = -\Delta\rho^{chiral}\sin\theta\cos\theta \text{ and } \rho_{xx}^{AMR} = \rho_\perp - \Delta\rho^{chiral}\cos^2\theta, \tag{5}$$

where the key parameter $\Delta\rho^{chiral} = \rho_\perp - \rho_{//}$ quantifies the chiral anomaly-induced resistivity anisotropy between perpendicular ($\theta = 90°$) and parallel ($\theta = 0°$) field orientations[27]. This comprehensive approach not only isolates the topological contributions but also provides precise quantification of the chiral anomaly's role in these interconnected transport phenomena.

Following our rigorous signal extraction protocol, we successfully isolate the intrinsic PHE and AMR signatures. **Figure 4b-c** reveal these purified signals exhibiting characteristic 180º periodicity, with the AMR profile displaying a notable 45º phase shift relative to the PHE signal - a direct consequence of their respective $\cos^2\theta$ and $\sin\theta\cos\theta$ angular dependencies. The temperature evolution of these effects, shown in **Figure 4d-f**, demonstrates several key features including that signal amplitudes maximize at 2 K, the extracted resistivity anisotropy $\Delta\rho^{chiral}$ exhibits non-monotonic temperature dependence, and a distinct anomaly emerges at $T_N$, unequivocally linking the transport anisotropy to magnetic ordering and associated spin textures - a previously overlooked correlation in such systems.

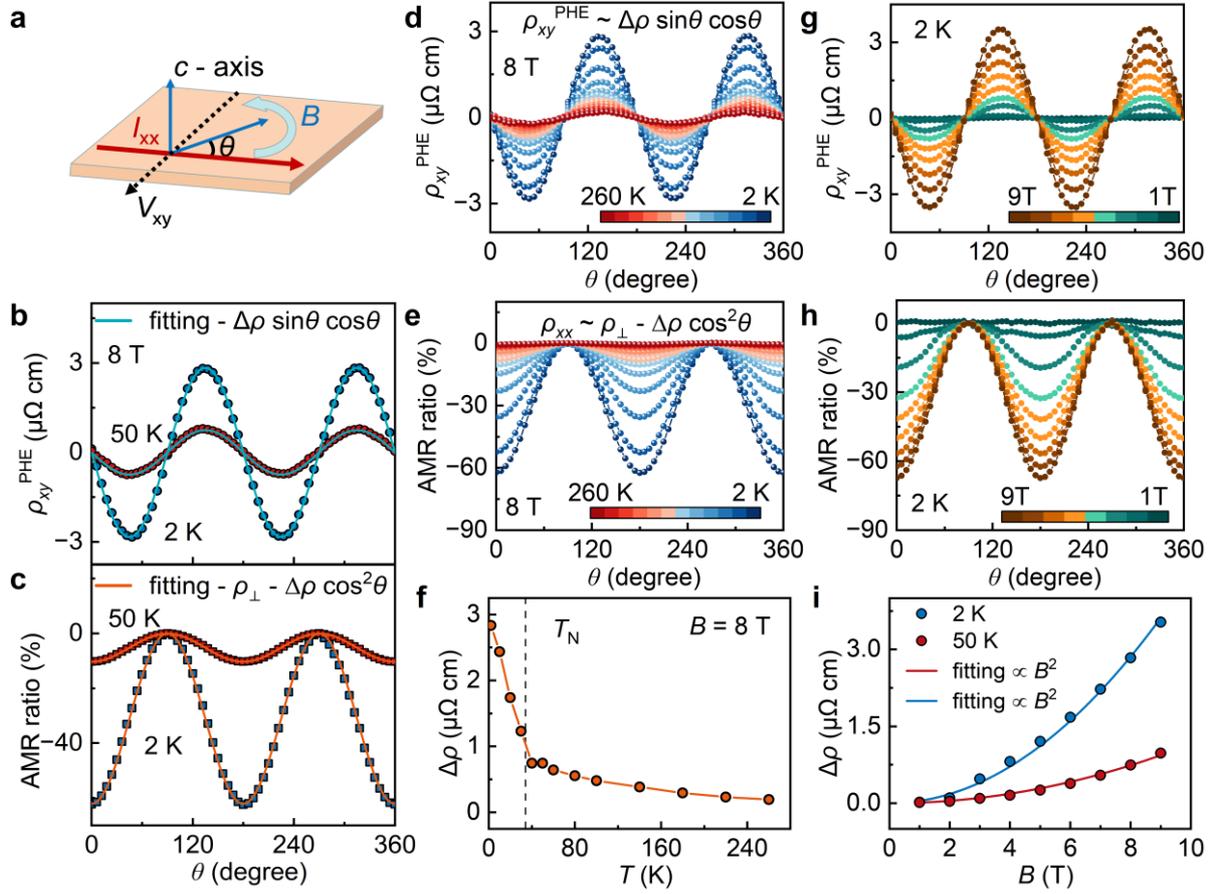

**Figure 4. AMR and PHE in EuAl$_2$Si$_2$ single crystals.** (a) Schematic illustration of the measurement geometry. The sample is rotated within the *ab*-plane while simultaneously recording both $I_{xx}$ and $V_{xy}$ voltage signals. Experimental PHE (b) and AMR (c) data (symbols) are fitted (solid curves) at $T$ = 2 K and 50 K under $B$ = 8 T. The excellent agreement between data and fits confirms the expected angular dependencies. (d,e) Angular evolution of (d) the off-diagonal PHE resistivity ($\rho_{xy}^{PHE}$) and (e) the diagonal AMR ratio across temperatures from 2-260 K at 8 T. The PHE signal exhibits characteristic $\sin\theta\cos\theta$ behavior, while the AMR follows a $\cos^2\theta$ dependence, as highlighted by the dashed guideline curves. (g,h) Field-dependent evolution of (g) $\rho_{xy}^{PHE}$ and (h) AMR ratio at 2 K for fields from 1-9 T. The PHE amplitude grows systematically with increasing field while maintaining its fundamental angular dependence. (f) Temperature dependence of the fitted PHE amplitude ($\Delta\rho$). The distinct kink at $T_N \sim 33.6$ K reflects the magnetic phase transition, with $\Delta\rho$ decreasing monotonically at higher temperatures. (i) Field dependence of $\Delta\rho$ at 2 K and 50 K. The quadratic scaling ($\Delta\rho$

∝ $B^2$, dashed lines) confirms the dominance of conventional PHE mechanisms in both the ordered and paramagnetic states.

Our field-dependent measurements (**Figure 4g-i**) at representative temperatures (2 K and 50 K) reveal quadratic *B*-dependence of $\Delta\rho^{chiral}$, reminiscent of conventional ferromagnets [28]. However, this similarity across distinct magnetic phases (FM *vs*. PM) with differing topological band structures complicates the unambiguous attribution to chiral anomaly effects. Theoretical framework suggests in the weak-field limit ($L_a \gg L_c$)[27]:

$$\Delta\rho^{chiral} \propto (L_c/L_a)^2, \qquad (6)$$

where $L_a = (D/\Gamma)B$ represents the magnetic length scale (*D*: diffusion coefficient, Γ: transport coefficient), and $L_c = (D\tau_c)^{1/2}$ denotes the chiral charge diffusion length ($\tau_c$: chiral charge scattering time). This relationship highlights $L_a$'s role in quantifying chiral anomaly strength through trivial-chiral charge coupling.

The observed nonlinear Hall response necessitates consideration of multiband transport dynamics. Our two-band model analysis yields the resistivity tensor for *B* //*x*:

$$\rho(B) = \begin{bmatrix} (\sigma_e + \sigma_h)^{-1} & 0 \\ 0 & (\sigma_e/\Delta_e + \sigma_h/\Delta_h)^{-1} \end{bmatrix}, \qquad (7)$$

where $\sigma_e = n_e\mu_e$ and $\sigma_h = n_h\mu_h$ represent electron and hole conductivities (*n*: carrier density, *μ*: mobility), with $\Delta_e = (1+\mu_e^2 B^2)$ and $\Delta_h = (1+\mu_h^2 B^2)$. While this model predicts *B*-independent longitudinal resistivity ($\rho_{xx}$), it reveals significant transverse conductivity ($\sigma_{yx}$) suppression - scaling as $1/B^2$ in high fields - arising from cyclotron motion of both carrier types[26,30,31]. The high-field limit yields $\rho_\perp \propto \frac{(\mu_e\mu_h)^2 B^2}{\sigma_h\mu_e^2 + \sigma_e\mu_h^2}$, resulting in quadratic-field-dependent $\Delta\rho^{chiral}$ without saturation, as $\rho_{//}$ remains field-invariant. This classical orbital mechanism can mimic chiral anomaly signatures[26,29,30], necessitating careful discrimination between these competing effects in our analysis.

Through systematic decomposition of longitudinal resistivity components, we elucidate the distinct mechanisms governing planar Hall effects across different magnetic regimes. Our analysis reveals that parallel ($\theta = 0°$) and orthogonal ($\theta = 90°$) field configurations selectively probe $\rho_{//}$ and $\rho_\perp$ respectively, with temperature-dependent characteristics clearly visible in symmetrized resistivity data (**Figure 5a** and **d**). At 2K, the system undergoes a field-induced AFM-to-FM transition via spin-flop reorientation, manifested in resistivity components that precisely track magnetic restructuring, where $\rho_\perp$ exhibits characteristic quadratic *B*-dependence in the FM state with low-field deviations (**Figure 5b**), while $\rho_{//}$ shows modest enhancement during spin-flop transition before saturating - behavior fully consistent with classical orbital effects. The contrasting 50 K behavior, where both components decrease with field before separating at higher fields (**Figure 5e**), directly contradicts chiral anomaly expectations. Parametric analysis of $\rho_{xy}$ versus $\rho_{xx}$ provides further mechanistic insight, with 2 K data showing distinctive "shock-wave" patterns (**Figure 5c**) from asymmetric $\rho_\perp$ enhancement that starkly contrast the isotropic expansion predicted for chiral-anomaly-dominated systems like $Na_3Bi$ and $GdPtBi$[32,33]. At 50 K, circular expansion patterns instead (**Figure 5f**) reflect field suppression of thermal spin fluctuations, with the longitudinal channel revealing enhanced dissipation information. These observations collectively demonstrate a temperature-dependent hierarchy of dominant mechanisms: thermal spin fluctuation suppression governs the PM regime, AFM→FM transition dynamics dominate the intermediate state, and classical orbital effects prevail in the FM state - all mediated through $Eu^{2+}$ spin scattering that creates characteristic unbalanced resistivity changes distinct from pure chiral anomaly signatures.

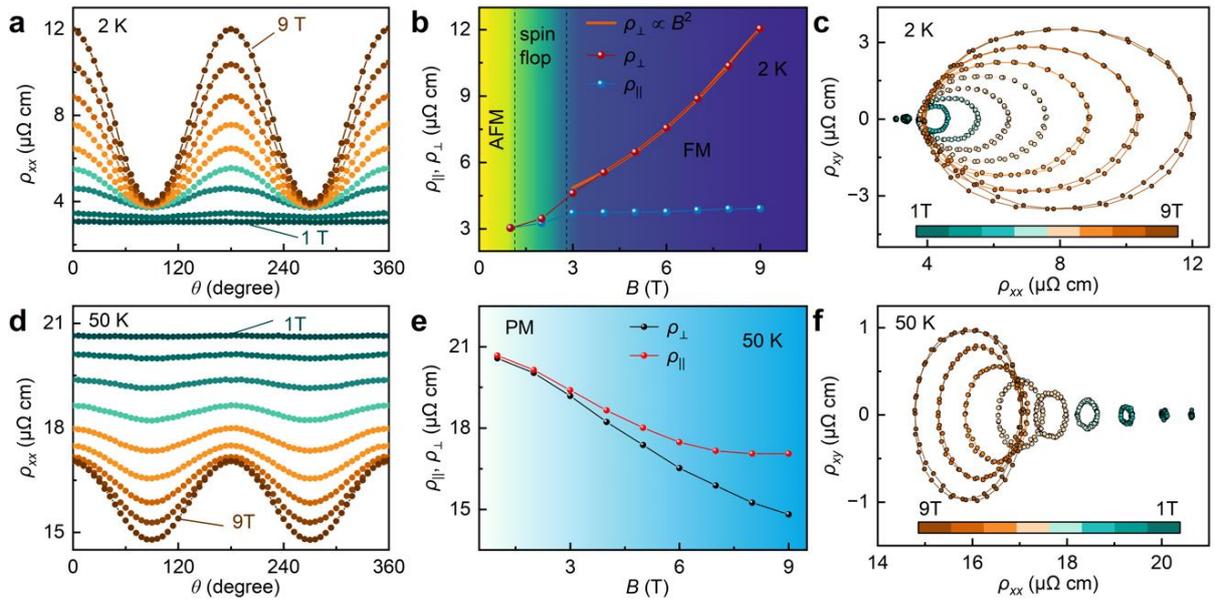

**Figure 5. Origin and evolution of planar Hall effect in EuAl₂Si₂.** (a,d) Temperature-dependent angular magnetoresistance profiles reveal distinct transport regimes. At 2 K (a), the AMR exhibits characteristic field evolution through the spin-flop transition, while the 50 K data (d) demonstrate PM behavior with modified angular dependence. (b) Field-dependent resistivity components ($\rho_\perp$ in red and $\rho_{//}$ in blue) at 2 K track the magnetic phase transition (background shading). The excellent quadratic fit to $\rho_\perp$ (orange curve) contrasts with the field-independent $\rho_{//}$, confirming classical orbital origin. The phase diagram highlights the AFM-FM transition via spin-flop reorientation. (e) High-temperature (50 K) resistivity components both decrease with field before separating at high fields, deviating from low-T behavior. This temperature crossover reflects the changing dominance of scattering mechanisms. (c,f) Parametric PHE trajectories provide crucial mechanistic fingerprints: FM state (2 K, c) shows characteristic "shock-wave" expansion from asymmetric $\rho_\perp$ enhancement, PM state (50 K, f) exhibits circular expansion patterns indicating spin fluctuation suppression. The contrasting geometries directly reflect the changing balance between orbital and spin-dependent transport processes.

## 3. Conclusion

Through comprehensive investigation of single-crystalline $EuAl_2Si_2$—a material uniquely hosting an AFM axion insulator to FM Weyl semimetal transition—we elucidate key electronic and magnetic properties governing the PHE. Magnetotransport measurements reveal a pronounced PHE with distinct origins across different magnetic regimes: in the PM state, it arises primarily from field-suppressed thermal spin fluctuations, while in the FM phase, classical orbital effects dominate. Although the predicted chiral anomaly signature remains elusive, this systematic separation and identification of competing mechanisms is critical for deconvoluting the PHE signal—a fundamental step toward its reliable application in next-generation spintronic memory, high-precision magnetic field sensors, and topological quantum devices. By providing a material platform where spin texture, magnetic order, and band topology are electrically tunable, our work establishes $EuAl_2Si_2$ as a paradigm-shifting system for engineering tailored PHE responses and probing emergent quantum phenomena for functional applications.

## 4. Experimental Section

*Crystal growth and chracterizations*: High-purity $EuAl_2Si_2$ single crystals were successfully synthesized via the self-flux method similar as that described previously [8]. The resulting crystals exhibited a stable, black metallic luster with no observable degradation or phase transitions under ambient conditions. The chemical composition of the $EuAl_2Si_2$ crystals was verified using scanning electron microscopy (SEM) coupled with energy-dispersive X-ray spectroscopy (EDS). Structural analysis was performed via powder X-ray diffraction (PXRD) on the (00*l*) crystallographic planes using a Bruker D8 Venture diffractometer (Cu Kα radiation, λ = 1.5418 Å). For higher-resolution crystallographic data, single-crystal X-ray diffraction (SXRD) was conducted on a Bruker D8 diffractometer equipped with Mo Kα radiation (λ = 0.71073 Å) at room temperature.

*Magnetization and magnetotransport measurements*: Magnetic characterization was carried out using a Quantum Design MPMS-3 magnetometer. Temperature-dependent magnetization $M(T)$ was measured under a 1000 Oe field applied both parallel ($H \mathbin{/\mkern-6mu/} ab$) and perpendicular ($H \perp ab$) to the *ab*-plane, with data collected in zero-field-cooled (ZFC) and field-cooled (FC) modes. Isothermal magnetization curves were acquired across a broad field range (±7 T) at multiple temperatures. Electrical transport properties were investigated via the standard four-probe method in a Quantum Design DynaCool PPMS. High-field magnetotransport measurements were performed at the Steady High Magnetic Field Facility (High Magnetic Field Laboratory, Chinese Academy of Sciences, Hefei).

*MFM* measurements: The EuAl$_2$Si$_2$ single crystal was cleaved in air to expose a pristine (001) surface. The cleavage quality was verified at room temperature using both optical microscopy and atomic force microscop, confirming a flat and uncontaminated surface. All MFM measurements were performed using a cryogenic MFM system under ambient pressure across various temperatures and applied magnetic fields. The MFM images were acquired over a 10 $\mu$m × 10 $\mu$m scan area. To mitigate electrostatic interference, the sample surface was coated with a conductive capping layer consisting of 2 nm Ti followed by 10 nm Au.

**Acknowledgements**

The authors acknowledge the National Key R&D Program of China (Grants No. 2023YFA1406100, 2024YFA1408400 and 2022YFA1403000). Y.F.G. acknowledges the open research funds of Beijing National Laboratory for Condensed Matter Physics (2023BNLCMPKF002) and Henan Key Laboratory of Aeronautical Materials and Application Technology (ZHKF-250103). W.X. thanks the support by the Shanghai Sailing Program (23YF1426900) and the National Natural Science Foundation of China (Grants No. 12404186). W.B.W. acknowledges the finial support by the National Natural Science Foundation of China (Grant No. 12374161). The authors also thank the support from Analytical Instrumentation Center (#SPST-AIC10112914) and the Double First-Class Initiative Fund of ShanghaiTech University.

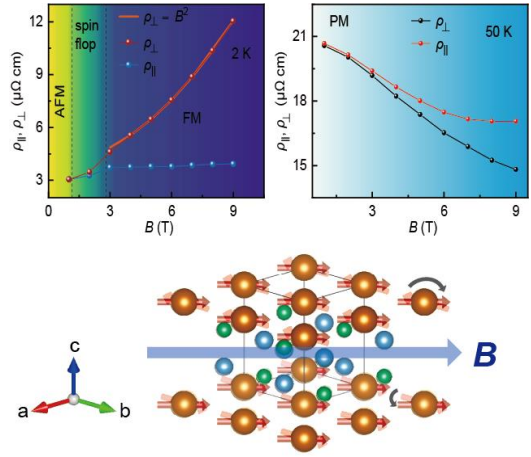